\documentclass[
journal=jacsat, 
manuscript=article]{achemso}

\usepackage[version=3]{mhchem}
\usepackage{graphicx}
\usepackage{subcaption}
\usepackage{bigints}
\usepackage{natbib}

\usepackage[nottoc]{tocbibind} 
\usepackage[english]{babel}
\usepackage[utf8]{inputenc}
\author{Ranveer Kumar Singh}
\affiliation[ Indian Institute of Science Education and Research Bhopal, Indore Bypass road, Bhauri, Bhopal Madhya Pradesh,India -462066.]
{Department of Mathematics\\Indian Institute of Science Education and Research Bhopal  \\Indore Bypass road, Bhauri, Bhopal Madhya Pradesh, India - 462066.\\}
\email{ranveer@iiserb.ac.in}

\title[\texttt{achemso} demonstration]
{On the classical dynamics of charged particle in special class of spatially non-uniform magnetic field}

\begin{document}

\begin{abstract}
Motion of a charged particle in uniform magnetic field has been studied in detail, classically as well as quantum mechanically. However, classical dynamics of a charged particle in non-uniform magnetic field is solvable only for some specific cases. We present in this paper, a general integral equation for some specific class of non-uniform magnetic field and its solutions for some of them. We also examine the supersymmetry of Hamiltonians in exponentially decaying magnetic field with radial dependence and conclude that this kind of non-uniformity breaks supersymmetry.
\end{abstract}
\begin{center}
    
\end{center}
\section{
    1.\hspace{0.5cm}Introduction}
It is known that a charged particle in a uniform magnetic field exhibits circular motion with specified frequency of revolution, called cyclotron frequency, and specified radius which depends on the charge and mass of particle and magnitude magnetic field [1]. Quantum mechanically, a charged particle in uniform magnetic field exhibits quantised energy levels, called Landau levels [2]. 

In classical arena, non-uniform magnetic fields of different kinds have different effects on dynamics of the charged particle. For instance, a charged particle in a magnetic field with non zero gradient undergo Grad B drift [3]. Charged particle moving along curved magnetic field experience centrifugal force perpendicular to magnetic field. Hence charged particles experience  drift in their motion. This kind of drift is called curvature drift [3]. There are several applications of these effects. One of the most important application is magnetic mirrors, which are used to confine plasmas [4]. The motion of an electron in a magnetic field of constant gradient has been analyzed by Seymour \textit{et al.}, where he has derived the $x$ and $y$ coordinates of electron's trajectory in terms of elliptic integrals [5]. The non-uniform magnetic field required for an electron to exhibit trochoidal trajectory has been calculated and it turns out to be a function of $x$ coordinate only [6]. The quantum mechanical treatment of charged particle in a class of non-uniform magnetic field has been studied using Isospectral Hamiltonian and Supersymmetric quantum mechanics [7]. This analysis gives the same Landau Level spectrum [8] as in case of uniform magnetic field. Although classical trajectory of a charged particle cannot be solved for the most general non-uniform magnetic field \textit{i.e,} non-uniform in all the three coordinates but it can still be solved for some classes of non-uniform magnetic field. It should be noted that throughout the paper, only spatially varying magnetic field is considered without time dependence. We use Landau gauge to restrict vector potential to just one component for constant magnetic field along $z$ direction. To inculcate non uniformity in magnetic field, we introduce some function in vector potential. Using elementary classical mechanics, we obtain an integral equation which can be solved to get the $x$ and $y$ coordinates of the particle's trajectory. Lastly, we examine the sypersymmetry structure of non-uniform Hamiltonians and observe that exponentially decaying magnetic field with radial dependence breaks supersymmetry.
\section{2.\hspace{0.5cm}Charged Particle in Uniform Magnetic Field}
First, we use the quadrature to compute the trajectory of a charged particle in a uniform magnetic field, which is known to be circular. Let us assume a constant magnetic field in $z$ direction \textit{i.e.} $\vec{B}=B\hat k$. Vector potential for constant magnetic field is given by, 
\begin{equation}
    \vec{A}=-\frac{1}{2}(\vec{r}\times \vec{B})
\label{Eq.1}    
\end{equation}
Direct calculation for the above magnetic field gives $A_x=-yB/2$ and $A_y=xB/2$ and the $z$ component is 0. We can choose Landau gauge to reduce $A$ to one component such that either $A_x=-yB$ or $A_y=xB$. Let us take $A_x=-yB$. Thus Lagrangian for the system can be written as,
\begin{center}
$\mathcal{L}=\sum_{i}(\frac{1}{2}m\dot q_i^2+qA\cdot\dot q_i)$
\end{center}
where the symbols have their usual meaning. Here, we can assume no motion in $z$ direction. So expanding the Lagrangian gives the following expression.
\begin{equation}
    \mathcal{L}=\frac{1}{2}m(\dot x^2+\dot y^2)-qyB\dot x
\label{Eq.2} 
\end{equation}
From Eq. \ref{Eq.2}, it is evident that $x$ is a cyclic coordinate so $p_x=\frac{\partial \mathcal{L}}{\partial \dot x}$ is conserved [1]. Thus we have $p_x=\frac{\partial \mathcal{L}}{\partial \dot x}=m\dot x-qBy=c$ or 
\begin{equation}
    \dot x=k_1+k_2y
\label{Eq.3} 
\end{equation}
where $k_1=c/m$ and $k_2=qB/m$. Since Lagrangian has no explicit time dependence, thus the total energy of the system is also a constant of motion. We can get the  Hamiltonian by using the following expression,
\begin{equation}
    \mathcal{H}=\sum_i(p_i\dot q_i-\mathcal{L})
\label{Eq.4} 
\end{equation} 
    where $p_i=\frac{\partial \mathcal{L}}{\partial \dot q_i}$. After putting $\mathcal{L}$ in Eq. \ref{Eq.4}, we have,
\begin{equation}
    \mathcal{H}=\frac{1}{2}m(\dot x^2+\dot y^2)
\label{Eq.5} 
\end{equation}
From Eq. \ref{Eq.3} and Eq. \ref{Eq.5} we get,
\begin{equation}
    \dot y=\pm \sqrt{k_3-k_1^2-k_2^2y^2-2k_1k_2y}
\label{Eq.6} 
\end{equation}
where $k_3=2\mathcal{H}/m$. Thus we have,
\begin{equation}
    \pm\frac{dy}{\sqrt{\alpha+\beta y+\gamma y^2}}=dt
\label{Eq.7} 
\end{equation}
where $\alpha=k_3-k_1^2$ , $\beta=-2k_1k_2$ and $\gamma=-k_2^2$. Integrating both sides with appropriate limits we get (one can check in integral table),
\begin{equation}
    \frac{1}{\sqrt{-\gamma}}\cos^{-1}[-(\beta+2\gamma y)/\sqrt{q}]=\pm t
\label{Eq.8} 
\end{equation}
with  $q=\beta^2-4\alpha\gamma$. Putting all the values and after simplifying we get,
\begin{equation}
y=\frac{\sqrt{k_3}}{k_2}\cos k_2t-\frac{k_1}{k_2}  \label{Eq.9}  \end{equation}
From Eq. \ref{Eq.3} we can integrate and calculate $x$ to get :
\begin{equation}
x=\frac{\sqrt{k_3}}{k_2}\sin k_2t
\label{Eq.10}
\end{equation}
Eqs. \ref{Eq.9} and Eq. \ref{Eq.10} define a circle (as can be checked easily by squaring and adding) with defined frequency of revolution, called cyclotron frequency $\omega=k_2=qB/m$. The radius of the orbit can be  calculated from Eq. \ref{Eq.9} and Eq. \ref{Eq.10}. Thus we get, $r=\frac{\sqrt{k_3}}{k_2}=mv/qB$ by noting that if energy of the system is conserved then $\mathcal{H}=mv^2/2$ where $v$ is the initial velocity of the particle. 
 \section{3.\hspace{0.5cm}Charged particle in a non-uniform magnetic field }
We now treat the general case of non-uniform magnetic field. We introduce a function depending on $y$ in the Landau gauge as $A_x=-yBf(y)$ which gives us non-uniform magnetic field. The magnetic field can be calculated easily using $B=\nabla\times A$ which gives,
\begin{equation}
\vec{B}=(yBf'(y)+Bf(y))\hat k
\label{Eq.11}
\end{equation}
where $f'(y)=\frac{df(y)}{dy}$. Eq.\ref{Eq.11} represents a special kind non-uniform magnetic field. If we specify $f(y)$ we obtain different classes of non-uniform magnetic field. We can again, assume no motion along $z$ axis.
Lagrangian for this case can be written as,
\begin{equation}
    \mathcal{L}=\frac{1}{2}m(\dot x^2+\dot y^2)-qyBf(y)\dot x
\label{Eq.12} 
\end{equation}
Considering symmetry of Lagrangian along $x$ we have $p_x=\frac{\partial \mathcal{L}}{\partial \dot x}=m\dot x-qByf(y)=c$ or :
\begin{equation}
    \dot x=k_1+k_2yf(y)
\label{Eq.13} 
\end{equation}
where $k_1=c/m$ and $k_2=qB/m$. Hamiltonian for the system can be calculated using Eq. \ref{Eq.4} and simple calculation gives Eq. \ref{Eq.5}. Here the Hamiltonian is a constant of motion as the Lagrangian has no explicit time dependence. From Eqs. \ref{Eq.5} and \ref{Eq.13} we have,
\begin{equation}
    \dot y=\pm\sqrt{k_3-k_1^2-k_2^2y^2f^2(y)-2k_1k_2yf(y)}
\label{Eq.14} 
\end{equation}
Eq. \ref{Eq.14} gives us the desired integral equation which can be solved for different $f(y)$ and in turn for different classes of non-uniform magnetic field to get $y$ as a function of $t$. This $y$ can be substituted in Eq. \ref{Eq.13} to get $x$ as a function of $t$. These two equations define the trajectory of the particle. The integral equation is :
\begin{equation}
 \boxed {\int \frac{dy}{\sqrt{k_3-k_1^2-2k_1k_2 yf(y)-k_2^2y^2f^2(y)}}=\pm\int dt +K } 
\label{Eq.15} 
\end{equation}
where $K$ is the constant of integration.
\section{3.1\hspace{0.5cm}Special Cases}
We can check that Eq. \ref{Eq.15} yields correct results for some special cases.\\
{\bfseries Case 1:} $f(y)$=1\\
In this case we get constant magnetic field along $z$ direction as can be checked by putting $f(y)=1$ in Eq. \ref{Eq.11}. Putting $f(y)=1$ in Eq. \ref{Eq.15} gives back Eq. \ref{Eq.7} which we have already solved. Thus we get circular trajectory for constant magnetic field.\\ 
{\bfseries Case 2:} $f(y)=1/y$\\
For this case, We can calculate that $\vec{B}=0$. This means particle must go undeviated \textit{i.e,} the trajectory should be a straight line. Putting $f(y)=1$, Eq. \ref{Eq.15} gives,
\begin{center}
$\int \frac{dy}{\sqrt{k_3-k_1^2-2k_1k_2-k_2^2}}=\pm t +K$
\end{center}
The denominator is just a constant (say $a$). So we get,
\begin{center}
$y=at+K'$ where $K'=aK$
\end{center}
It can be noted that both the signs $+$ and $-$ give the same  trajectory. We can calculate $x$ from Eq. \ref{Eq.13} which comes out to be :
\begin{center}
$x=bt+D$ where $b=k_1+k_2$ and $D$ is a constant of integration.
\end{center}
$x$ and $y$ indeed define a straight line. We can check that $\dot x^2+\dot y^2=v^2$ where $v$ is the initial velocity of the particle.

We can extent the same procedure for a non-uniform magnetic field in $x$ coordinate. In that case we use the Landau gauge in which $A_x=0=A_z$ and $A_y=xBf(x)$. Magnetic field in this case takes the following form,
\begin{equation}
\vec{B}=(xBf'(x)+Bf(x))\hat k
\label{Eq.16} 
\end{equation}
Accordingly Lagrangian assumes the form given by,
\begin{equation}
  \mathcal{L}=\frac{1}{2}m(\dot x^2+\dot y^2)+qxBf(x)\dot x  
\label{Eq.17} 
\end{equation}
Now $y$ is a cyclic coordinate so that $p_y=\frac{\partial \mathcal{L}}{\partial \dot y}=m\dot x+qBxf(x)=c$ or 
\begin{equation}
    \dot y=k_1-k_2xf(x)
\label{Eq.18} 
\end{equation}
where $k_1=c/m$ and $k_2=qB/m$. Hamiltonian of the system remains the same. Invoking energy conservation and using Eqs. \ref{Eq.17} and \ref{Eq.5} we obtain,
\begin{equation}
    \dot x=\pm\sqrt{k_3-k_1^2-k_2^2x^2f^2(x)+2k_1k_2xf(x)}
\label{Eq.19} 
\end{equation}
This gives us another integral equation which can be solved to obtain particle's $x$ coordinate as a function of time which can be substituted in Eq. \ref{Eq.17} to obtain particle's $y$ coordinate. Thus we can obtain particle's trajectory. The integral equation is given as :
\begin{equation}
 \boxed {\int \frac{dx}{\sqrt{k_3-k_1^2+2k_1k_2 xf(x)-k_2^2x^2f^2(x)}}=\pm\int dt +K } 
\label{Eq.20} 
\end{equation}
where $K$ is the constant of integration. The above cases give us the same result as it  should be.
\section{
    4.\hspace{0.5cm}Exponentially decaying Magnetic field}
Although Eq. \ref{Eq.15} can be solved for different kinds of non uniform magnetic fields, but let's consider an exponentially decaying magnetic field. Let the magnetic field be given by,
\begin{equation}
    \vec{B}=Be^{-y}\hat k
\label{Eq.21} 
\end{equation}
where $B$ is a constant. From Eq. \ref{Eq.11} we can solve for $f(y)$. Thus we have,
\begin{center}
$Be^{-y}=yBf'(y)+Bf(y)$
\end{center}
This is an ordinary differential equation. Solution to this is given by,
\begin{equation}
    f(y)=\frac{c}{y}-\frac{e^{-y}}{ y}
\label{Eq.22} 
\end{equation}
We can fix $c=1$ so that,
\begin{equation}
    f(y)=\frac{1}{y}(1-e^{-y})
\label{Eq.23} 
\end{equation}
magnetic field still remains the same. Putting this in Eq. \ref{Eq.15} gives,
\begin{equation}
  {\int \frac{dy}{\sqrt{k_3-k_1^2-2k_1k_2 (1-e^{-y})-k_2^2(1-e^{-y})^2}}=\pm\int dt +K } 
\label{Eq.24} 
\end{equation}
Let us put $a=k_3-k_1^2$ , $b=k_2^2$ and $c=2k_1k_2$ then the integral has the solution given by :
\begin{equation}
  \dfrac{\arcsin\left(\frac{2\left(c+b-a\right)\mathrm{e}^{y}-c-2b}{\sqrt{c^2+4ab}}\right)}{(\sqrt{c+b-a})}+K'=\pm t+K
\label{Eq.25} 
\end{equation}
To make equations look simpler let's put $c+b-a=\alpha^2$ , $c+2b=\beta$ and by calculation  $c^2+4ab=2k_2\sqrt{k_3}$. Constants of integration can be manipulated such that $K'=K$ so the solution after rearranging the terms looks as,
\begin{equation}
  y=\log\bigg(\frac{\sqrt{k_3}k_2}{\alpha^2}\sin (\pm\alpha t)+\frac{\beta}{2\alpha^2}\bigg)
\label{Eq.26} 
\end{equation}
From Eq. (13) we can calculate $x$ as,
\begin{equation}
  x=\int \bigg(k_1+k_2-\frac{k_2}{l\sin(\pm\alpha t)+m}\bigg)dt
\label{Eq.27} 
\end{equation}
where $l=\frac{\sqrt{k_3}k_2}{\alpha^2}$ , $m=\frac{\beta}{2\alpha^2}$. Solution of this equation is,
\begin{equation}
  x=(k_1+k_2)t-k_2\dfrac{2\arctan\left(\frac{m\tan\left(\frac{\alpha t}{2}\right)\pm l}{\sqrt{m^2-l^2}}\right)}{\alpha\sqrt{m^2-l^2}}
\label{Eq.28} 
\end{equation}
Simple calculation gives,
\begin{equation}
  x=(k_1+k_2)t-2\arctan\left(\frac{(m/l)\tan\left(\frac{\alpha t}{2}\right)\pm 1}{\sqrt{(m/l)^2-1}}\right)
\label{Eq.29} 
\end{equation}
in terms of original constants we have,
\begin{equation}
  x=(k_1+k_2)t-2\arctan\left(\frac{(k_1+k_2)\tan\left(\frac{\alpha t}{2}\right)\pm \sqrt{k_3}}{\sqrt{(k_1+k_2)^2-k_3}}\right)
\label{Eq.30} 
\end{equation}
Eqs. \ref{Eq.26} and \ref{Eq.30} define the trajectory of the particle. It is worth noting that if we have a particle with zero energy in magnetic field then it will remain stationary since magnetic field does no work. In this case this is indeed true. If we take zero energy then $k_3=0$ and a bit of calculation shows that $x=0$ and $y$=log($k_2$)=constant. Thus the particle remains stationary on point (0, log($k_2$)). 

\begin{figure}[h]
\centering
\includegraphics[width=0.5\textwidth]{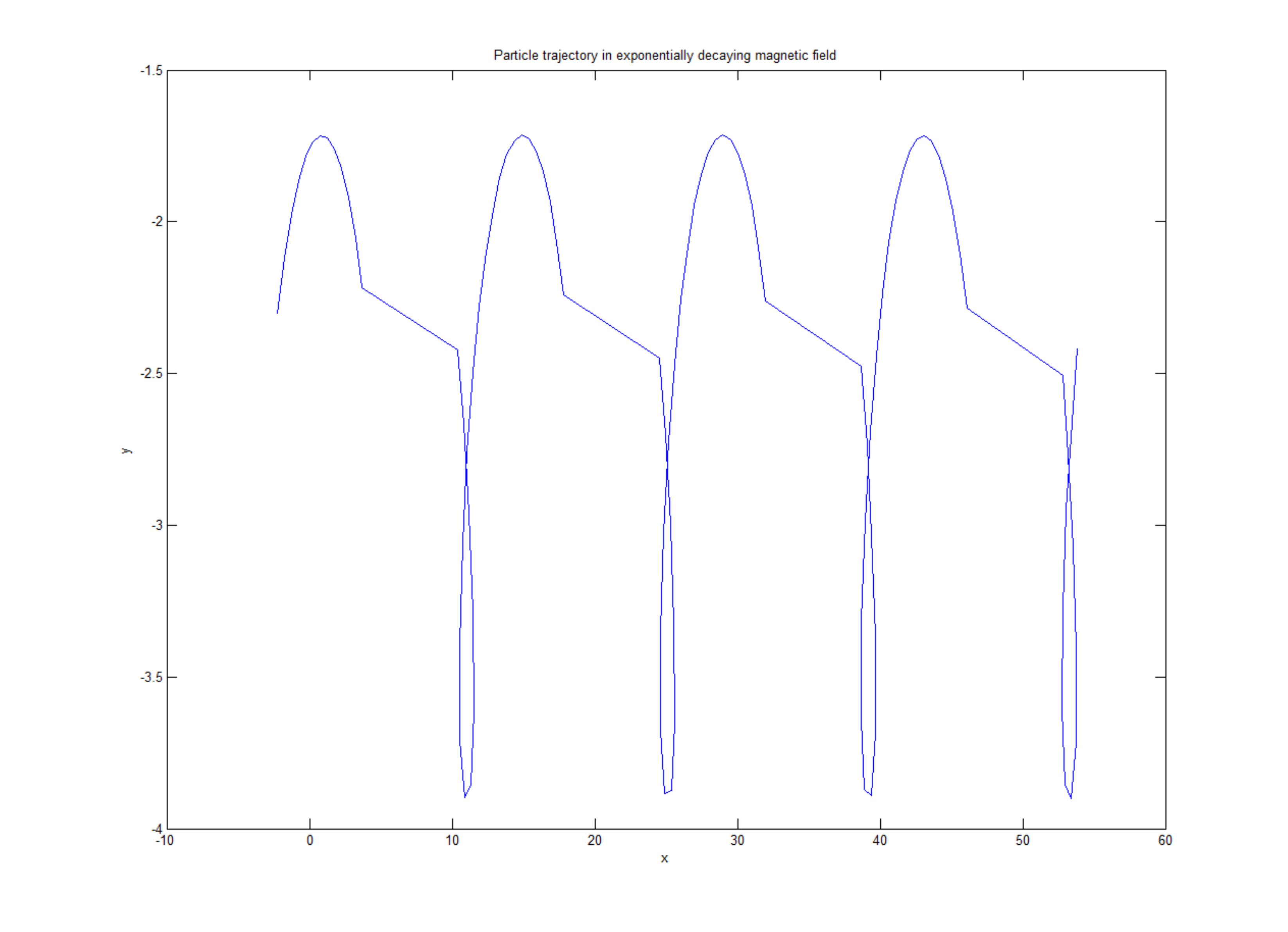}
\caption{Particle's trajectory in exponentially decaying magnetic field $\vec{B}=Be^{-y}\hat k$}
\label{fig.1}
\end{figure}

Now to analyze particle's trajectory in this field let's  assume the following :
$k_3=1$ , $k_2=0.1$ (we have taken specific charge of the particle to be $1$ and assumed a strong magnetic field of $0.1$T!) ;$k_1+k_2=1.118$. This gives us,
\begin{center}
$x=1.12t-2\tan^{-1}(2.24\tan(t/4)+2)$;\\
$y=\log(0.08\sin(t/2)+0.1)$
\end{center}
We plot these parametric equation for $t\in (0,50)$. In this Fig. \ref{fig.1} and the figures that follow, the horizontal axis is the x-axis and the vertical axis is the y-axis.
The particle shows periodic motion as can be inferred from the trajectory. We plot trajectory curves for several values of constants. There are several forms of $f(y)$ for which solution to Eq. \ref{Eq.15} exists and thus such class of non-uniform magnetic field can be analyzed easily. 
\begin{figure}[h]
\begin{subfigure}{.5\textwidth}
  \centering
  \includegraphics[width=.8\linewidth]{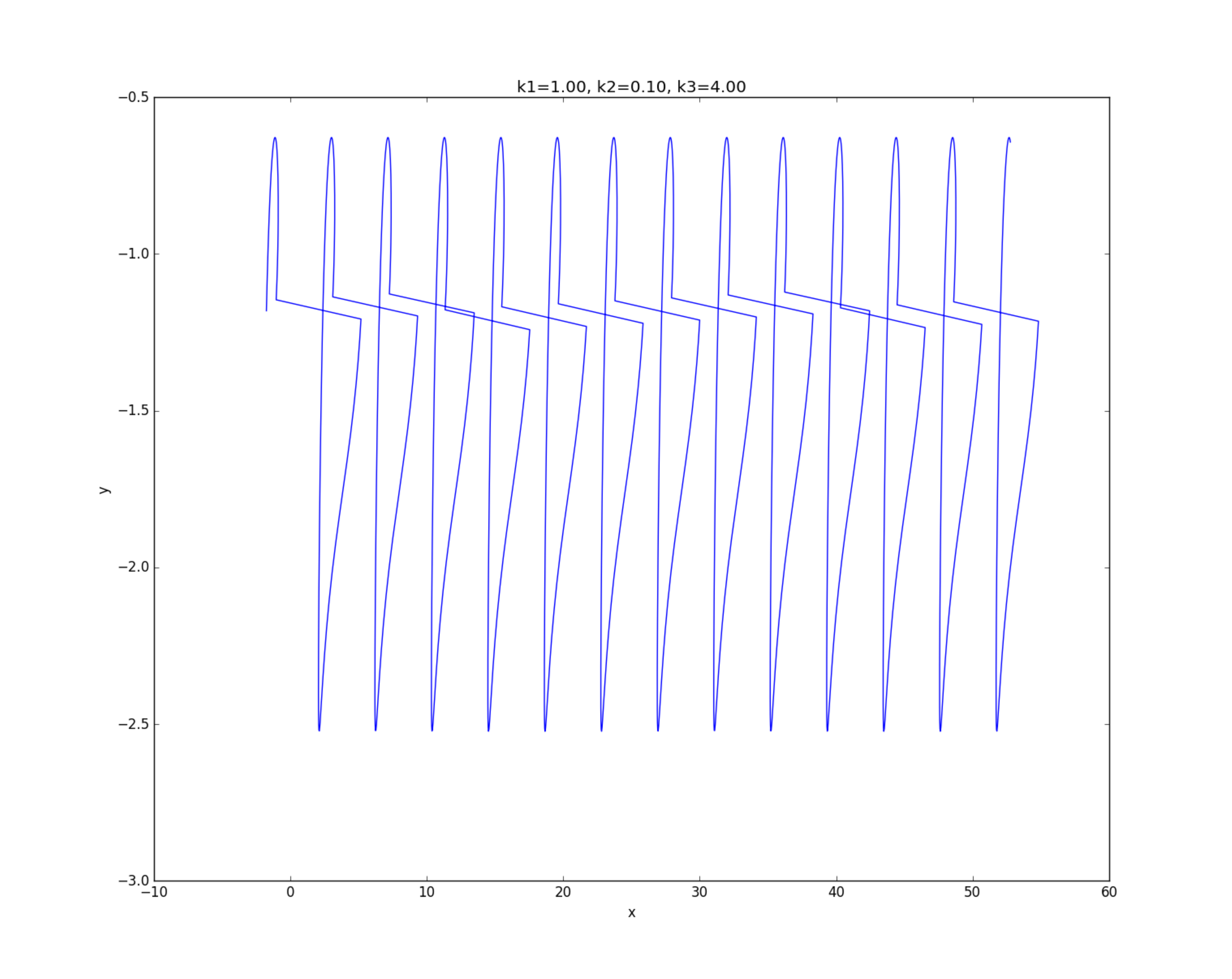}
  \caption{$k_1=1 , k_2=0.1 , k_3=4$}
 \label{fig.2a} 
\end{subfigure}%
\begin{subfigure}{.5\textwidth}
  \centering
  \includegraphics[width=.8\linewidth]{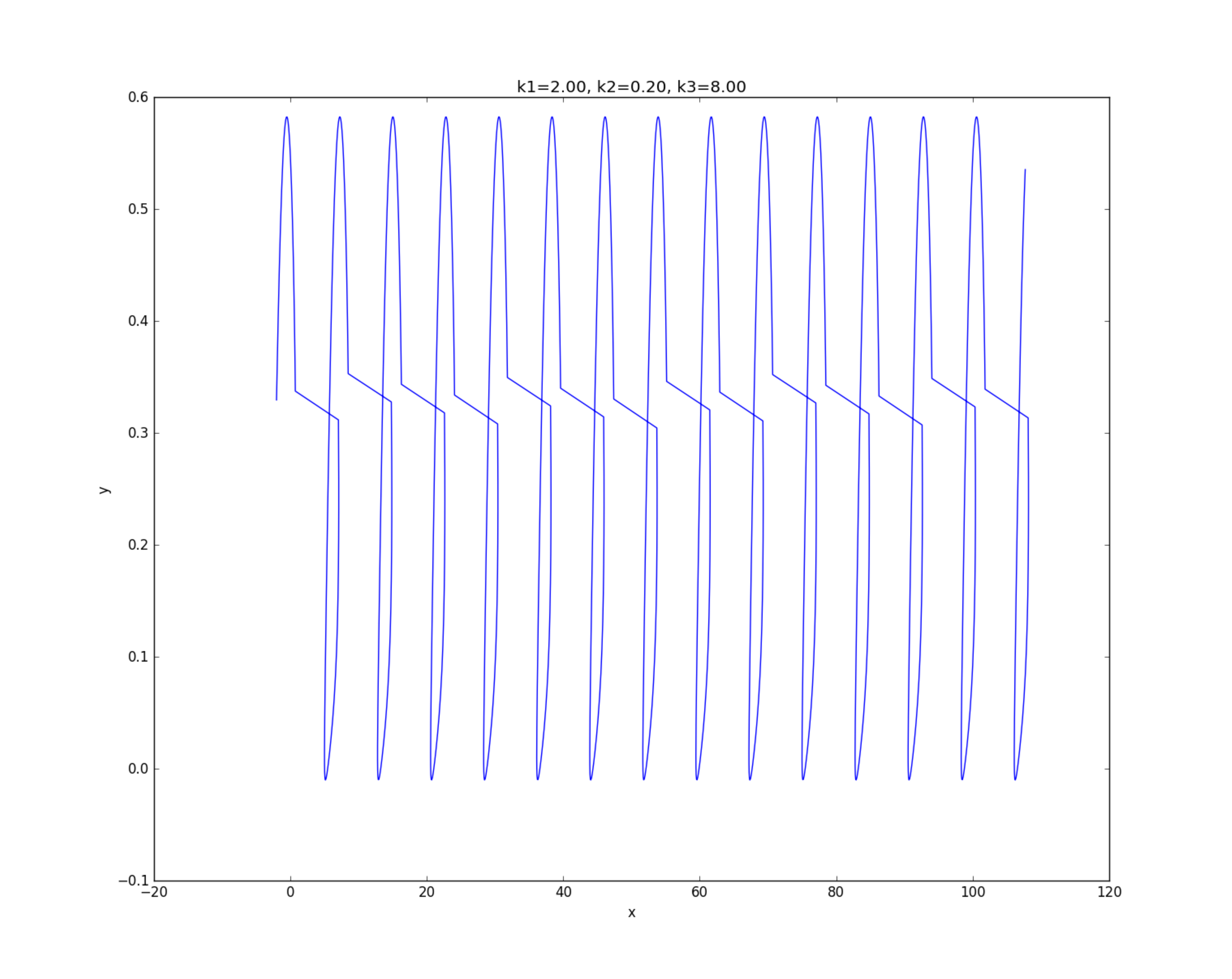}
  \caption{$k_1=2 , k_2=0.2 , k_3=8$}
 \label{fig.2b} 
\end{subfigure}
\begin{subfigure}{.5\textwidth}
  \centering
  \includegraphics[width=.8\linewidth]{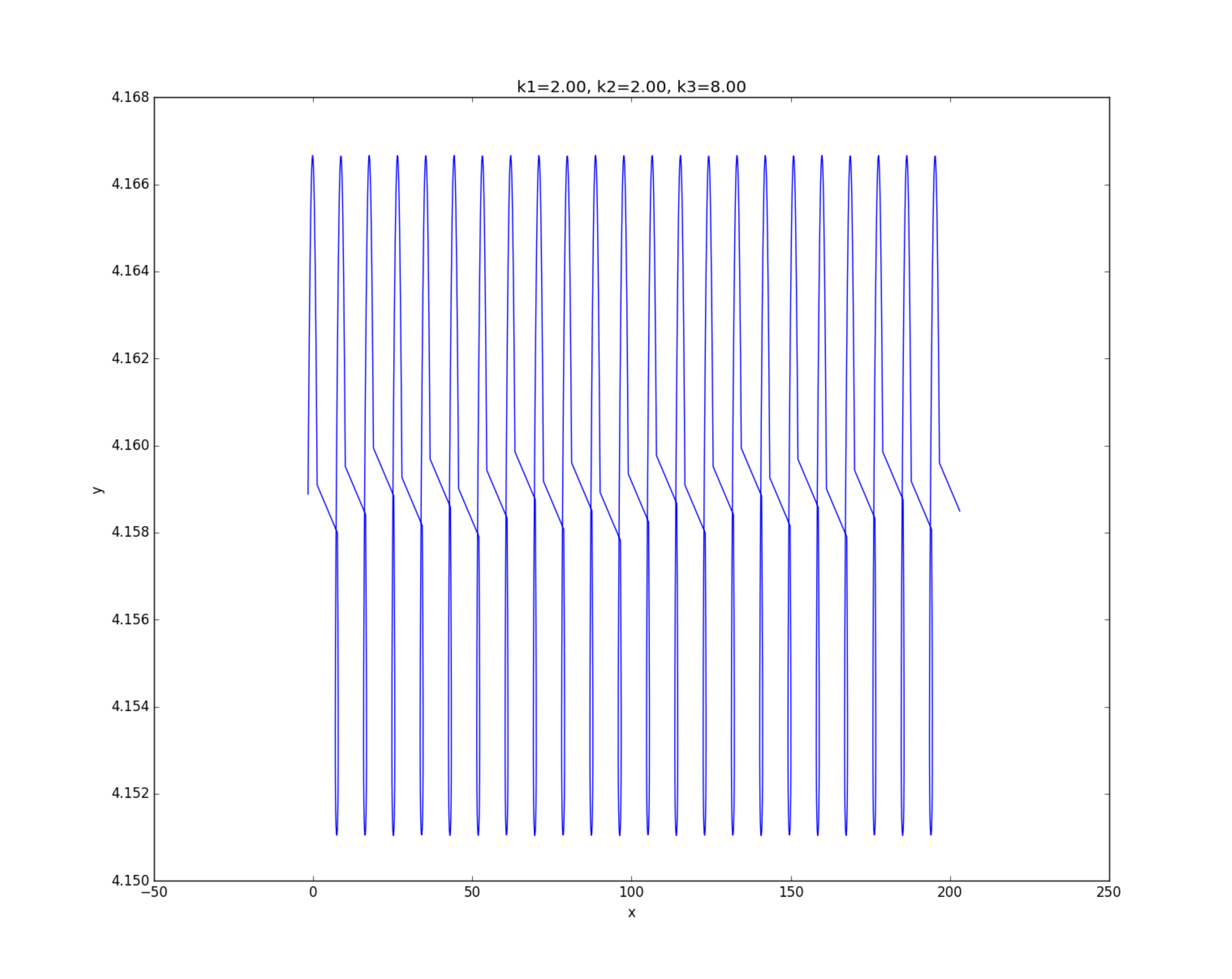}
  \caption{$k_1=2 , k_2=2 , k_3=8$}
\label{fig.2c}  
\end{subfigure}\begin{subfigure}{.5\textwidth}
  \centering
  \includegraphics[width=.8\linewidth]{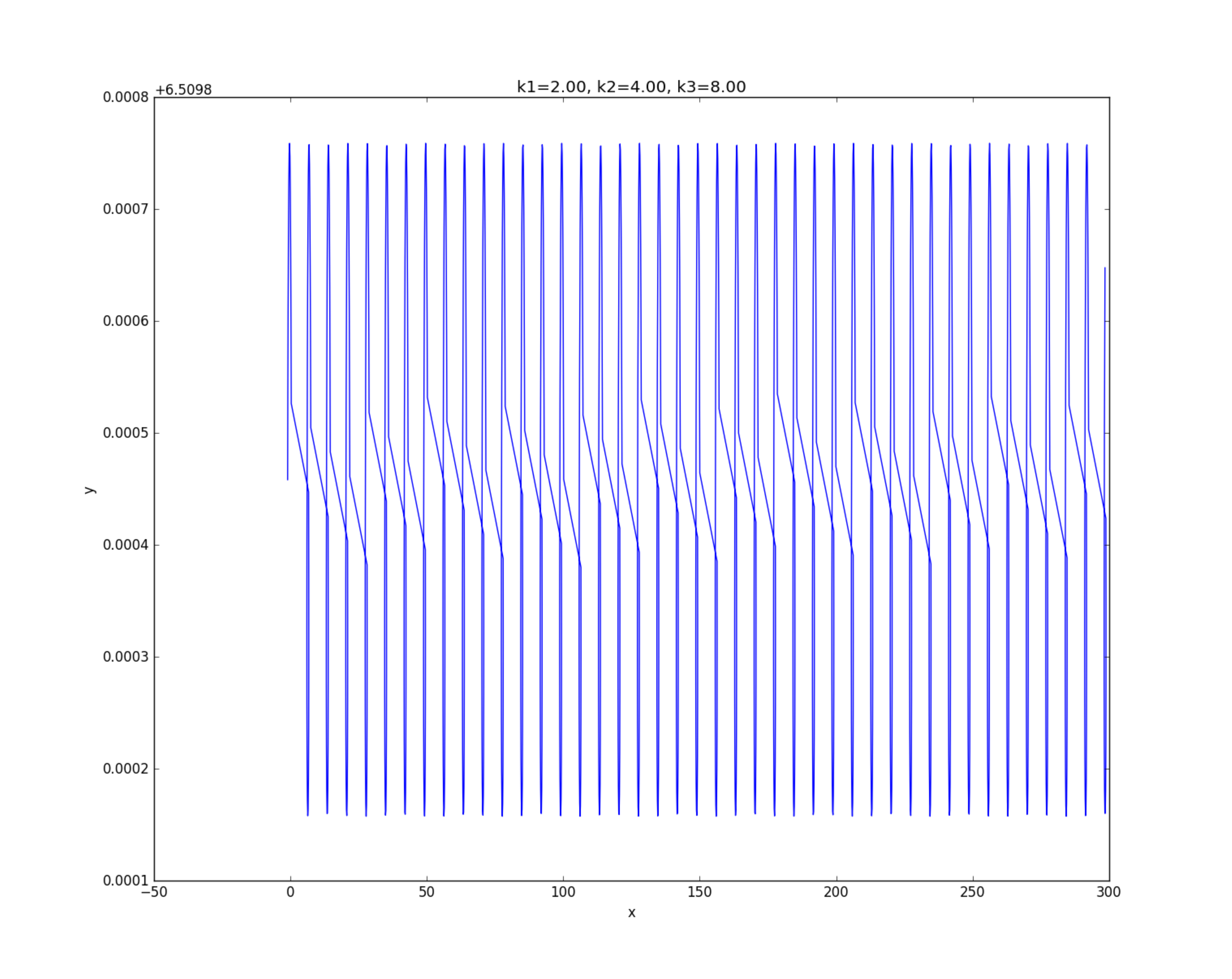}
  \caption{$k_1=2 , k_2=4 , k_3=8$}
 \label{fig.2d}  
\end{subfigure}
\caption{Plots of trajectory curves for different values of constants}

\end{figure}
\section{5.\hspace{0.5cm}Supersymmetry in Uniform and Non-Uniform \\Magnetic Field}
Now we turn to the quantum mechanical treatment of the problem. The Pauli-Hamiltonian for a charged particle moving in two dimensions in a magnetic field [9] is given by, 
\begin{equation}
    2H = (p_x+A_x)^2+(p_y+A_y)^2+(\nabla \times A)_z\sigma_z 
\label{Eq.31} 
\end{equation}
where $\sigma_z$ is the Pauli z matrix. We have used natural units with $\hbar=1=m$. Suppose, we choose $A_x=-Byf(r)$ and $A_y=B_xf(r)$ where $r=\sqrt{x^2+y^2}$ then Eq. \ref{Eq.31} has the following form,
\begin{equation}
    2H=-\Big(\frac{d^2}{dx^2}+\frac{d^2}{dy^2}\Big)+B^2r^2f^2-2BfL_z+(2Bf+Brf'(r))\sigma_z
\label{Eq.32} 
\end{equation}
where $L_z$ is the z-component of the orbital angular momentum operator. We use cylindrical coordinates $(r, \phi)$ to solve the corresponding Schr$\ddot{o}$dinger equation. The wave function $\psi(r, \phi)$ can be factored as,
\begin{equation}
    \psi(r,\phi)=R(r)e^{im\phi}
\label{Eq.33} 
\end{equation}
where $m=0,\pm 1,\pm 2,\pm 3, \dots$ are the eigenvalues of the operator $L_z$. On substituting Eq.\ref{Eq.33} into Eq. \ref{Eq.32} we obtain the following equation,
\begin{equation}
    \frac{d^2R}{dr^2}+\frac{1}{r}\frac{dR}{dr}-\Big[B^2r^2f^2+\frac{m^2}{r^2}+2Bmf+ (2Bf+ Bpf'(r)\sigma_z)\Big]R(r)=-2ER(r)
\label{Eq.34} 
\end{equation}
If we further substitute $R(r)=\sqrt{r}A(r)$ into Eq. \ref{Eq.34} and choose the lower eigenvalue of $\sigma_z$, we obtain :
\begin{equation}
    \Bigg[-\frac{d^2}{dr^2}+\Bigg(B^2r^2f^2-2Bf+2Bmf-Brf'(r)+\frac{m^2-\frac{1}{4}}{r^2}\Bigg)\Bigg]A(r)=2EA(r)  
\label{Eq.35} 
\end{equation}
We can write the left hand side in the form $a^{\dagger}a$ where 
\begin{equation}
    a= \frac{d}{dr}+Brf-\frac{|m|+\frac{1}{2}}{r}
\label{Eq.36} 
\end{equation}
For $m\le0$, the decomposition holds and thus we have that $E_0\ge0$ [10]. $E_0=0$ occurs if and only if the solution of the equation $a\psi_0(r)=0$ \textit{i.e,}
\begin{equation}
    \Bigg[\frac{d}{dr}+Brf-\frac{|m|+\frac{1}{2}}{r}\Bigg]\psi_0(r)=0
\label{Eq.37} 
\end{equation}
is square integrable [10]. In that case  Supersymmetry (SUSY) remains unbroken. Now consider the following cases :
\begin{enumerate}
    \item \textbf{Uniform magnetic field}\\
    If we choose $f(r)=1$ above, we end up having a uniform magnetic field. In this case, we can solve Eq. \ref{Eq.37} to get :
    \begin{equation}
        \frac{d\psi_0(r)}{dr}=\Bigg(\frac{|m|+\frac{1}{2}}{r}-Br\Bigg)\psi_0(r)
   \label{Eq.38} 
    \end{equation}
    \begin{equation}
        \frac{d\psi_0(r)}{\psi_0(r)}=\Bigg(\frac{|m|+\frac{1}{2}}{r}-Br\Bigg)dr
    \label{Eq.39} 
    \end{equation}
Solution to Eq. \ref{Eq.39} is given by 
\begin{equation}
    \psi_0(r)=N_0r^{|m|+\frac{1}{2}}exp\Big(-\frac{1}{2}Br^2\Big)
\label{Eq.40} 
\end{equation}
where $N_0$ is the normalization factor. It is easy to see that $\psi_0(r)$ is square integrable as the polynomial factor is dominated by the exponential and overall integral is convergent. Thus in this case the SUSY remains unbroken. It is also known that in uniform magnetic field, the energy spectrum is same as that of a harmonic oscillator oscillating with cyclotron frequency. 
\item \textbf{Non-Uniform magnetic field}\\
There are several forms of $f(r)$ for which SUSY remains unbroken. For instance the function $f(r)=\frac{(r-a)(r-b)}{r^2}$ keeps the SUSY unbroken. Now consider the function 
\begin{equation}
    f(r)=\frac{1}{r}(1-e^{-r})
\label{Eq.41} 
\end{equation}
For this function, first note that $\vec{B}=\nabla\times \vec{A}=(yBf'(y)+Bf(y))\hat k=Be^{-r}\hat k$. Thus for this function we obtain exponentially decaying magnetic field, this time with radial decay, which we considered in previous section for classical treatment. Now we deal with the peculiarity of this form. We can solve Eq. \ref{Eq.37} with this function and a bit of computation gives 
\begin{equation}
    \psi_0(r)=N_0r^{|m|+\frac{1}{2}}exp\big[-B(e^{-r}+r)\big]
\label{Eq.42} 
\end{equation}
Again, we can easily see that $\psi_0(r)$ is not square integrable as the integral diverges. Thus we have $E_0>0$ and thus \textit{SUSY is broken}. Energy spectrum still remains discretly quantised [10] and can be determined using 
\begin{equation}
    \bigintsss_0^{\infty}\sqrt{2m[E_n-W^2(r)]dr}=\Big(n+\frac{1}{2}\Big)\hbar\pi,  n=0,1,2,3,\dots
\label{Eq.43} 
\end{equation}
where $W(r)$ is given by
\begin{equation}
    W(r)=-\frac{\hbar}{\sqrt{2m}}\frac{\psi_0'(r)}{\psi_0(r)}
\label{Eq.44} 
\end{equation}
Thus we see that exponentially decaying magnetic field is really peculiar as it breaks supersymmetry. Further experiments may be designed to detect this kind of symmetry breaking.
\end{enumerate}
    
\section{6.\hspace{0.5cm}Discussion}
The analysis presented in this paper provides an important recipe to find the trajectory of a charged particle in spatially varying magnetic field for some special classes of non uniform magnetic field. The more general case of solving the trajectory still remains an open problem. Nonetheless the example presented provides a way to solve the trajectory in exponentially varying magnetic field. We also observe that exponentially decaying magnetic field shows special properties in the sense that the ground state of the isospectral partner of the non-uniform Hamiltonian with exponential non-uniformity has non-zero ground state energy and thus it breaks supersymmetry. Experiments may be designed to detect this feature of exponentially decaying magnetic field. We also note that the results of the paper may be used in various other cases. Physical situations where non-uniform magnetic field appears is that of the earth itself. New insights can be gained by studying the trajectory of charged particles released from the sun in the earth's magnetic field. Study of the van Allen belts can be done based on this theory.
\section{7.\hspace{0.5cm}Conclusion  }
In conclusion, here we have presented a method to analyze the motion of a charged particle in various classes of non-uniform magnetic field. We also presented an specific non-uniform magnetic field with peculiar properties classically as well as quantum mechanically. Although the integral equation presented do not have trivial solution for many forms of $f(y)$ or equivalently $f(x)$ but for such cases numerical integration can be used to find out the trajectory. It is observed that many forms have exact solution and thus it is useful to further study this method and generalize it to other forms.
\section{8.\hspace{0.5cm}Acknowledgements}
This work was carried out at Panjab University Chandigarh. The author is indebted to Prof. C. N. Kumar whose guidance helped to complete this work. The author also thanks Ms. Harneet Kaur, Dr. Amit Goyal, and Mr. Shivam Pal  for useful discussions.

\section{9.\hspace{0.5cm}References }

[1] Goldstein, H.; Poole, C. P. and Safko, J. L. \textit{ Classical Mechanics }(3rd ed.). Addison-wesley (2001)\\
\big[2\big] Landau, L. D. and Lifschitz, E. M.;\textit{ Quantum Mechanics: Non-relativistic Theory}. Course of Theoretical Physics. Vol. 3 (3rd ed. London: Pergamon Press). ISBN 0750635398 (1977)\\
\big[3\big] Baumjohann, W. and Treumann, R. \textit{Basic Space Plasma Physics}. ISBN 978-1-86094-079-8 (1997)\\
\big[4\big] Krall,N. \textit{Principals of Plasma Physics}. Page 267 (1973)\\
\big[5\big] Seymour,P.W - \textit{Aust. Jour. Phys.} 12; 309-14\\
\big[6\big] RF Mathams, R.F. \textit{Aust. Jour. Phys}. 17(4), 547 - 552 \\
\big[7\big] Cooper,F.; Khare,A.; and Sukhatme,U. \textit{“Supersymmetry in quantum mechanics"},\textit{ Phys.
Rep}. 251, 267-285, (1995)\\
\big[8\big] Khare,A. and Kumar,C.N. \textit{Mod. Phys. Lett.} A 8, 523-529 (1993)\\
\big[9\big] Khare,A. and Maharana,J. \textit{Nucl. Phys.} B224, 409 (1984)\\
\big[10\big] Cooper,F.; Khare,A.; and Sukhatme,U. \textit{Phys.
Rep} 251, 267-285, (1995)

\end{document}